\documentclass[
twocolumn,
prl,
showpacs,
amsmath,
amssymb,
superscriptaddress,
floatfix
]{revtex4}

\usepackage{bm}
\usepackage{dsfont}
\usepackage{graphicx}
\usepackage{pifont}
\usepackage{textcomp}
\usepackage{gensymb}
\usepackage{accents}
\usepackage{upgreek}
\usepackage{color}

\makeatletter
\def\NAT@bibsetnum#1{%
 \setlength{\topsep}{\z@}%
 \NATx@bibsetnum{#1}%
}%
\renewenvironment{thebibliography}[1]{%
 \NAT@thebibliography{#1}%
 \@clubpenalty\clubpenalty
 \let\@TBN@opr\present@bibnote
 \@FMN@list
}{%
 \@endnotesinbib
 \edef\@currentlabel{\arabic{NAT@ctr}}%
 \NAT@endthebibliography
 \global\let\auto@bib\@empty
}
\newcommand*{\supplementarystart}{%
  \close@column@grid%
  \clearpage%
  \onecolumngrid%
  \setcounter{enumiv}{0} 
  \setcounter{equation}{0} 
  \setcounter{figure}{0} 
  \setcounter{table}{0} 
  \setcounter{page}{1}
  \c@secnumdepth=4
  \renewcommand{\theequation}{s\arabic{equation}} 
  \renewcommand{\bibnumfmt}[1]{[s##1]} 
  \renewcommand{\@onlinecite}{s\citealp} 
  \renewcommand{\cite}[1]{{[}\onlinecite{##1}{]}}
  \renewcommand{\thefigure}{s\arabic{figure}}
  \renewcommand{\thetable}{s\Roman{table}}
  \renewcommand{\thepage}{s\arabic{page}}
}
\makeatother

\newcommand{\s}{\sum\limits}

\newcommand{\be}{\begin{equation}}
\newcommand{\e}{\end{equation}}
\newcommand{\beml}{\begin{subequations}}
\newcommand{\eml}{\end{subequations}}
\newcommand{\beq}{\begin{eqnarray}}
\newcommand{\eq}{\end{eqnarray}}
\newcommand{\ba}{\begin{array}}
\newcommand{\ea}{\end{array}}
\newcommand{\bpm}{\begin{pmatrix}}
\newcommand{\epm}{\end{pmatrix}}
\newcommand{\bc}{\begin{cases}}
\newcommand{\ec}{\end{cases}}
\newcommand{\lt}{\left}
\newcommand{\rt}{\right}
\newcommand{\n}{\nonumber}

\newcommand{\ep}{\varepsilon}

\newcommand{\bb}{\boldsymbol}
\newcommand{\h}{^\dagger}
\newcommand{\0}{^{\phantom{\dagger}}}

\DeclareMathOperator{\im}{Im}

\begin{document}
	
\title{Magnon activation by hot electrons via non-quasiparticle states}
\author{S.\,Brener}
\affiliation{Radboud University, Institute for Molecules and Materials, NL-6525 AJ Nijmegen, The Netherlands}
\affiliation{I.\,Institut f\"ur Theoretische Physik, Universit\"at Hamburg, Jungiusstra{\ss}e 9, D-20355 Hamburg, Germany}
\author{B.\,Murzaliev}
\affiliation{Radboud University, Institute for Molecules and Materials, NL-6525 AJ Nijmegen, The Netherlands}
\author{M.\,Titov}
\affiliation{Radboud University, Institute for Molecules and Materials, NL-6525 AJ Nijmegen, The Netherlands}
\author{M.\,I.\,Katsnelson}
\affiliation{Radboud University, Institute for Molecules and Materials, NL-6525 AJ Nijmegen, The Netherlands}
	
\begin{abstract}
We consider the situation when a femtosecond laser pulse creates a hot electron state in half-metallic ferromagnet (e.\,g. ferromagnetic semiconductor) on a picosecond timescale but do not act directly on localized spin system. We show that the energy and magnetic moment transfer from hot itinerant electrons to localized spins is facilitated by the so-called non-quasiparticle states, which are the scattering states of a magnon and spin-majority electron. The magnon distribution is described by a quantum kinetic equation that we derive using the Keldysh diagram technique. In a typical ferromagnetic semiconductor such as EuO magnons remain essentially in non-equilibrium on a scale of the order of microsecond after the laser pulse.
\end{abstract}
	
\maketitle
	
Recently, a huge experimental progress has been reached in ultrafast laser-induced manipulation of magnetic properties of materials, with a time scale from picoseconds to femtoseconds \cite{Bigot2005,Koopmans2005,Koopmans2016,Rasing2005,Rasing2011,Rasing2014,Melnikov2005,Melnikov2008,RasingRMP}. This opens up a way to theoretical \cite{Secchi2013,Mentink2015,Itin2015,Secchi2016} as well as experimental \cite{FeO} studies of magnetic interactions out-of-equilibrium. Certainly we are just at the very beginning of a long journey and much more theoretical effort is needed to get microscopic understanding of magnetic phenomena at this ultra-short time scale.
	
It is especially interesting and important that lasers can act selectively at a given electron subsystem, e.\,g., only at localized 4f or itinerant 5d electrons in rare-earth metals \cite{Melnikov2005,Melnikov2008} and semiconductors \cite{Rasing2014}. The problem is relevant for spintronics \cite{Sarma2004} since lasers can make spintronic devices ultrafast \cite{RasingRMP}. In this work we consider the process of energy and spin transfer from non-equilibrium itinerant-electron system to the subsystem of localized magnetic moments within the framework of s-d(f) exchange model \cite{Vonsovsky1946,Zener1951,Vonsovsky1974}. Below we focus on a particular example of degenerate ferromagnetic semiconductors \cite{Nagaev1983,Nagaev2001} with the Fermi energy which is much smaller than the exchange splitting of the conduction band. The conductivity of these materials is fully dominated by electrons with a certain spin projection. Thus, the degenerate magnetic semiconductors can be considered as a special case of half-metallic ferromagnets \cite{Groot1983,Katsnelson2008} that include also such metallic systems as Heusler alloys (e.\,g. NiMnSb and PtMnSb), CrO$_2$ and related compounds with europium chalcogenides, e.\,g. EuO, and chromium-based spinels, e.\,g. CdCr$_2$Se$_4$, being prototype examples \cite{Nagaev1983}. Our theory also applies to colossal magnetoresistance materials such as (La,Ca)MnO$_3$ \cite{Nagaev2001} and dilute magnetic semiconductors such as Ga$_{1-x}$Mn$_x$As \cite{Dietl2014}.
	
Although equilibrium properties of ferromagnetic semicondustors and half-metallic ferromagnets in general are fairly well understood \cite{Nagaev1983,Nagaev2001,Katsnelson2008,Dietl2014}, a little is known on strongly non-equilibrium dynamics of such systems on time scales determined by electron-magnon interactions. Laser-induced magnon activation in these materials by femtosecond laser pulses is an important process that may provide new spintronic device functionalities. In this Letter we focus on the magnon activation in a half-metal ferromagnet mediated by the presence of the so-called non-quasiparticle states.
	
The non-quasiparticle states in a half-metal magnet have been first considered by Hertz and Edwards \cite{HE1973} in 1973 and extensively studied in the eighties by Irkhin and Katsnelson \cite{IK}. Some implications of non-quasiparticle states in real materials has been analyzed in Ref.~\cite{Chioncel2003} (for a review, see Ref.~\onlinecite{Katsnelson2008}). The non-quasiparticle states may be thought of as collective excitations of a spin-majority electron and a magnon. The existence of such states results in a finite (power law) density of states of minority electrons at and above the Fermi energy, even though the excitation of spin-minority electrons in the corresponding non-interacting model is forbidden. The non-quasiparticle states make a drastic impact on magnetic, spectroscopic and thermodynamic properties of half-metal ferromagnets at finite temperatures \cite{Katsnelson2008,IK}. Spin depolarization near the Fermi energy induced by the non-quasiparticle states was experimentally detected in Co$_2$MnSi \cite{Chioncel2008}. We stress that the existence of non-quasiparticle states is a correlation-induced effect. Together with few other phenomena, like Kondo effect \cite{Hewson}, it shows that the standard quasiparticle band theory may be insufficient even for qualitative understanding of properties of magnetic materials.
	
The principal goal of this paper is to demonstrate that the activation dynamics of magnons in the half metal is determined primarily by the non-quasiparticle states. A laser pulse creates a hot quasi-equilibrium itinerant electron state in the half-metal on a picosecond time scale, when the number of magnons is still negligible. We investigate how the energy and magnetic moment are transferred from the hot electrons to localized spins. It is demonstrated that this process is slow and creates an essentially non-equilibrium time-dependent magnon distribution that is described by a quantum kinetic equation derived below.
	
To be more specific we consider a ferromagnet described by the $s$-$d$ exchange model
\be
\label{starting}
H=H_s+H_d-I\!\sum_{\bb{k}\bb{q}\alpha\beta}\!\!\!\bb{S}_{\bb{q}}\,c\h_{\bb{k}+\bb{q} \alpha}\bb{\sigma}_{\alpha\beta}c\0_{\bb{k}\beta},
\e
where $H_s=\sum_{\bb{k}\sigma}\ep_{\bb{k}\sigma}c\h_{\bb{k}\sigma}c\0_{\bb{k}\sigma}$ describes itinerant (s) electrons characterized by the isotropic dispersion $\ep_{\bb{k},\sigma}=\ep_{\bb{k}}+\delta_{\sigma\downarrow}\Delta$ with $\ep_{\bb{k}}=k^2/2m$ and the effective electron mass $m$, the strength of $s$-$d$ exchange $I$ defines the Stoner gap $\Delta=2IS$, and $H_d$ describes core magnons with quadratic dispersion $\omega_{\bb{q}}=Dq^2$ that is cut off at the Debay frequency $\omega_D=D q_D^2$.
	
For a typical ferromagnetic semiconductor, such as doped EuO \cite{Nagaev1983}, one finds the gap $\Delta \simeq 1$\,eV and the Fermi energy $\ep_\textrm{F}\simeq 0.3$\,eV. In the absence of $s$-$d$ interactions these values would correspond to perfectly polarized itinerant electron system. Taking EuO as an example we get $mD \simeq 10^{-3} \ll 1$ and $S=7/2$ (parameters are taken from the book \onlinecite{Nagaev1983}). The energy $W=q_D^2/2m \simeq 10$\,eV roughly corresponds to a band width that is the largest energy scale in the problem.
	
The condition $2S\gg 1$ justifies the Holstein-Primakov transformation that reduces Eq.~(\ref{starting}) to a simpler model $H = H_s+H_d+H_{sd}+H_{mm}$, with $H_d =\sum_{\bb{q}}\omega_{\bb{q}}b^{\dagger}_{\bb{q}}b\0_{\bb{q}}$ and
\be
\label{hamil}
H_{sd}= -\frac{\Delta}{\sqrt{2S}} \s_{\bb{k},\bb{q}}
\lt(c^{\dagger}_{\bb{k+q}\downarrow}c\0_{\bb{k}\uparrow}b\0_{\bb{q}}+ \textrm{h.c.}\rt),\vspace*{-6pt}
\e
where $b\h_{\bb{q}}$ ($b\0_{\bb{q}}$) are magnon creation (annihilation) operators for the momentum $\bb{q}$. The term $H_{mm}$ describes the magnon-magnon interaction that we discuss later.
	
Below we focus on magnon activation that takes place on much longer time scales than the electron-electron relaxation time. It is legitimate to assume that the $s$-electron subsystem is heated up to the temperature $T$ within a picosecond after the laser pulse, while the transfer of energy to localized spins is a much slower process.
	
From a naive point of view the half-metal system defined by Eqs.~(\ref{starting},\ref{hamil}) cannot support low-energy magnons. Indeed, at low temperatures all electrons belong to the majority band. A magnon excitation, which relies upon the presence of a minority spin, would, on the first glance, require a huge excitation energy of the order of $\Delta-\ep_\textrm{F}$. This logic, however, fails due to the presence of $s$-$d$ interactions. Indeed, the minority spin can always be virtually excited in order to enable a magnon excitation at low energies. This problem was analyzed in detail in Ref.~\cite{IK} where the general concept of non-quasiparticle states has been introduced (based on the previous studies of the Hubbard model \cite{HE1973}). Those are the states that involve a virtual excitation of the minority spin that does not belong to the mass shell. The non-quasiparticle state can also be reinterpreted as a bound state of a magnon and a majority electron. Mathematically, non-quasiparticle states are seen as branch cuts (not poles) of the electron Green's function. At zero temperature they lead to the presence of a finite density of states of minority spins for all energies above the Fermi level.
	
In order to illustrate the concept of non-quasiparticle states more quantitatively one should compute the imaginary part of the retarded self-energy, $\im{\Sigma^{R}_{\downarrow}(\ep,\bb{k})}$, for the minority spin component of the equilibrium electron Green's function. Such an analysis to the lowest order in the electron-magnon interaction (and also to the small parameter $1/2S$) has been undertaken in Refs.~\cite{IK}. The corresponding result is given by
\begin{align}
\label{Gamma}
&\Gamma(\ep,\bb{k})=-2\im{\Sigma^{R}_{\downarrow}(\ep,\bb{k})}\\ \n
&= \frac{\Delta^2 V_D}{8\pi^2S}\int \!\!d^3\bb{q}\,(1+N_{\bb{q}}-f_{\bb{k}-\bb{q},\uparrow})\delta(\ep-\omega_{\bb{q}}-\ep_{\bb{k}-\bb{q},\uparrow}),
\end{align}
where $N_{\bb{q}}$ refers to the number of magnons with momentum $\bb{q}$, $f_{\bb{k},\uparrow}=f(\ep_{\bb{k},\uparrow})$ is the electron Fermi distribution $f(\ep)=\lt[\exp[(\ep-\ep_\textrm{F})/T]+1\rt]^{-1}$, and $V_D=3/4\pi q_D^3$ is the normalization volume that is given by the volume of a unit cell in the continuum limit.
	
At an initial time we let $N_{\bb{q}}=0$ in Eq.~(\ref{Gamma}), hence
\be
\label{Gammageneral}
\Gamma(\ep,\bb{k})=\frac{\Delta^2V_Dm}{8\pi S D k}\int_{\omega_-}^{\omega_+}\!\!\!\! d\omega\;\lt(1-f(\ep-\omega)\rt),
\e
where $\omega_\pm \equiv Dq_{\pm}^2$ with $q_{\pm}=|k\pm\sqrt{2m\ep}|$. (Here we used that $mD\ll 1$). The range of integration is dictated by the energy conservation law (represented by the delta-function in Eq.~(\ref{Gamma})) for $\omega_{\bb{q}} \ll \ep_\textrm{F}$.
	
It is instructive to analyze the behavior of $\Gamma(\ep,\bb{k})$ with temperature. It is easy to see that $\Gamma$ is exponentially suppressed for $T\ll \omega_-(\ep,\bb{k})$. Since $\omega_-$ itself is a function of $\ep$ and $\bb{k}$ this condition must be viewed as a self-consistent condition for $T\ll mD\ep_\textrm{F}\sim 3$\,K that effectively pins energy in Eq.~(\ref{Gammageneral}) to the mass shell of majority spin electrons. We do not analyze this temperature regime in detail due to its limited experimental significance.
	
For high temperature regime, $T\gg \omega_\textrm{D}$ we find that $T\gg \omega_+(\ep,\bb{k})$ for all momenta. In this case, the expression for $\Gamma(\ep,\bb{k})$ for energies near the Fermi surface and $T\gg D(k+k_\textrm{F})^2$ can be simplified to $\Gamma(\ep,\bb{k})= \Delta^2 mV_Dk_\textrm{F}\lt(1-f(\ep)\rt)/2\pi S$, where $k_\textrm{F}=\sqrt{2m\ep_\textrm{F}}$ is the Fermi momentum for majority electrons.
	
For intermediate temperatures, $mD\ep_\textrm{F}< T<\omega_\textrm{D}$, both cases $T<\omega_+$ and $T>\omega_+$ are realized for different values of momenta that prevents further analytical simplifications.
	
The hot electron distribution can be characterized by the non-equilibrium Green's functions $G^<$ and $G^>$ that are related to each other in temperature equilibrium
as $G^{>}_{\bb{k},\sigma}(\ep) = e^{(\ep-\ep_F)/T} G^{<}_{\bb{k},\sigma}(\ep)$. To the leading order in the parameter $1/2S$ we find
\be
\label{glesser}
G^<_{\bb{k},\downarrow}(\ep)=\frac{i\Gamma(\ep,\bb{k})f(\ep)}{(\ep_{\bb{k},\downarrow}-\ep)^2+\Gamma^2(\ep,\bb{k})/4},
\e
where $\Gamma(\ep,\bb{k})$ in the denominator of the Green's function is kept only to assure a formal numerical convergence of the integrals at the quasiparticle pole.
	
In what follows we consider electrons that are locally equilibrated due to phonons and/or electron-electron interactions and are characterized by a constant ``hot'' temperature $T$. We also assume that magnons are absent at the initial moment of time, $N_{\bb{q}}(t=0)=0$ and construct the kinetic equation for the number of magnons $N_{\bb{q}}(t)$. This equation can be further generalized by taking into account the dependence of electron temperature on time due to the heat transfer to the bath.  As the energy of the magnon subsystem is much lower than the energy of the electron subsystem we may always neglect the cooling of electrons due to energy transfer to magnons.
	
Let us now analyze the spin-flip rates, which are responsible for creation or annihilation of magnons. In ferromagnetic semiconductor with $T\ll\Delta-\ep_\textrm{F}$ the spin-flips take place only in the presence of non-quasiparticle states. Since the latter lack any dispersion, the operator $c\h_\downarrow c\0_\downarrow$ cannot be regarded as a quasiparticle number operator. Instead, the expectation value of such an operator has to be found using lesser or greater electron Green's functions considering energy and momentum as separate variables, i.\,e.\,beyond the mass shell.
	
The resulting kinetic equation reads
\be
\label{kinetic}
dN_{\bb{q}}/dt = \mathcal{I}_{sd}[N_{\bb{q}}],
\e
where the $s$-$d$ collision integral is given by
\begin{align}
&\mathcal{I}_{sd}=\frac{\Delta^2V_D}{2S}\!\!\int\!\!\frac{d\ep\, d^3\bb{k}}{(2\pi)^4}
\Big[(1+N_{\bb{q}})G^<_{\bb{k}+\bb{q}\downarrow}(\ep+\omega_{\bb{q}})G^>_{\bb{k}\uparrow}(\ep)\n\\
&\qquad\quad -N_{\bb{q}}G^>_{\bb{k}+\bb{q}\downarrow}(\ep+\omega_{\bb{q}})G^<_{\bb{k}\uparrow}(\ep)\Big].
\label{kineq}
\end{align}
The integration over energy $\ep$ in Eq.~(\ref{kineq}) is easily performed since the latter is pinned to the mass shell of majority electrons.  From Eq.~(\ref{glesser}) we obtain the $s$-$d$ collision rate as
\begin{align}
&\mathcal{I}_{sd}=\frac{\Delta^2V_D}{2S}\lt(1-\frac{N_{\bb{q}}}{n_{\bb{q}}}\rt)
\int \frac{d^3\bb{k}}{(2\pi)^3}(1-f(\ep_{\bb{k}}))f(\ep_{\bb{k}}+\omega_{\bb{q}})\n\\
&\frac{\Gamma(\ep_{\bb{k}}+\omega_{\bb{q}},\bb{k}+\bb{q})}{(\Delta+\ep_{\bb{k}+\bb{q}}-\ep_{\bb{k}})^2+\Gamma^2(\ep_{\bb{k}}+\omega_{\bb{q}},\bb{k}+\bb{q})/4},
\label{kineqsimpl}
\end{align}
where $n_{\bb{q}}=1/(e^{\omega_{\bb{q}}/T}-1)$ is the equilibrium distribution of magnons and the rate $\Gamma$ depends on $N_{\bb{q}}(t)$ (see Eq.~(\ref{Gamma})).
	
\begin{figure}[bt]
\includegraphics[width=\columnwidth]{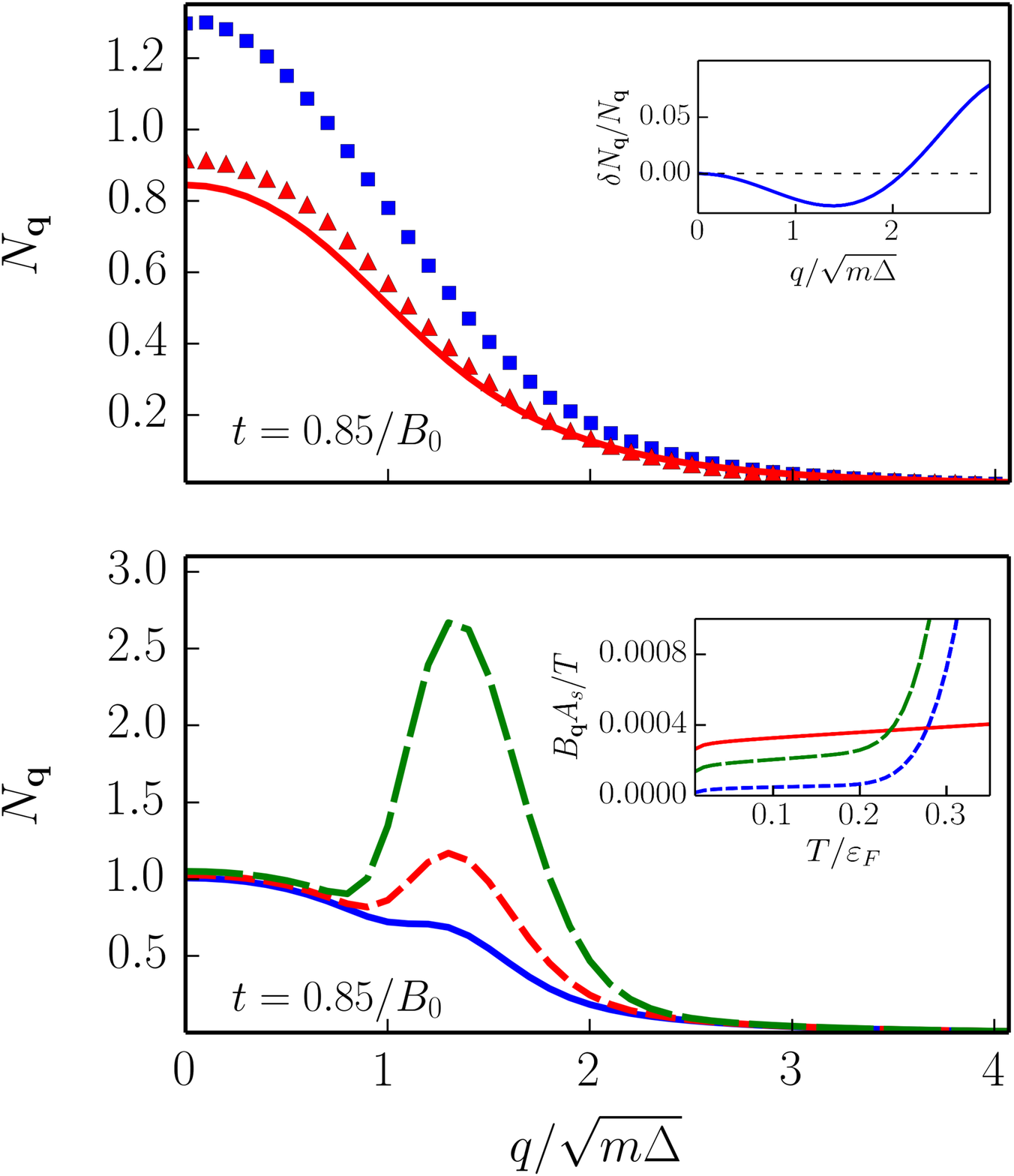}
\caption{The dependence of magnon distribution $N_{\bb{q}}$ on the modulus of the wave vector at the time $t\simeq B_0^{-1}$. The parameters relevant for EuO have been used. The upper panel illustrates the result for sufficiently low temperature $T=0.1\,\ep_F$. Square dots show the numerical solution to the kinetic equation (\ref{kinetic},\ref{kineqsimpl}). Triangular dots show the numerical solution of Eq.~(\ref{Equation0}), while the corresponding approximation (\ref{Bq}) is plotted with the solid line. The inset in the upper panel shows the relative correction $\delta N_{\bb{q}}/N_{\bb{q}}\simeq C_{\bb{q}} t^2/B_{\bb{q}}$ due to the magnon-magnon interactions computed from Eq.~(\ref{mm}). The lower panel illustrates the numerical solution of Eqs.~(\ref{kinetic},\ref{kineqsimpl}) for higher temperatures: $T=0.2 \ep_F$ (solid line), $T=0.22 \ep_F$ (short dashed line), and $T=0.25 \ep_F$ (long dashed line). In this regime the quasiparticle contribution $\delta B_{\bb{q}}$ to the magnon excitation rate is clearly visible. The inset shows the dependence of the rate on temperature for $q/\sqrt{m\Delta}=0$ (solid line), $1$ (long dashed line), and $2$ (short dashed line).}
\label{fig:Nq}
\end{figure}

The numerical solution of the kinetic equation (\ref{kinetic}) is illustrated in Fig.~\ref{fig:Nq} for experimentally relevant parameters.  For small times we can disregard the effect of finite $N_{\bb{q}}$ in $\Gamma$. In this limit the kinetic equation is given by
\be
\label{Equation0}
dN_{\bb{q}}/dt= \big(1-N_{\bb{q}}/n_{\bb{q}}\big)B_{\bb{q}},
\e
where we introduced a momentum-dependent growth rate $B_{\bb{q}}$ for the number of magnons (see the upper panel of Fig.~\ref{fig:Nq} for the illustration). For the case of intermediate temperatures, $\omega_\textrm{D} \ll T\ll\ep_\textrm{F}$, one can estimate from Eq.~(\ref{kineqsimpl}) that
\be
\label{Bq}
B_{\bb{q}}^{\mathrm{nq}}\simeq
\frac{T \lt(\Delta/W\rt)^2 \ep_F/W}{A_S [(1+\ep_{\bb{q}}/\Delta)^2-4\ep_{\bb{q}}\ep_\textrm{F}/\Delta^2]},
\e
where $A_S=\pi(4\pi S)^2(8\pi/3)^2 \approx 10^5$ for $S=7/2$. For $T \approx 100$\,K and for other parameters typical for EuO, the relevant time scale $1/B_0$ is of the order of a microsecond, which implies that the magnon distribution $N_{\bb{q}}(t)$ is strongly non-equilibrium at all relevant times.

In addition to Eq.~(\ref{Bq}) there exists also a quasiparticle contribution $\delta B_{\bb{q}}$ from the pole in Eq.~(\ref{kineqsimpl}),
\be
\label{Bqqp}
\delta B_{\bb{q}}\simeq\frac{3T(\Delta/W)^{3/2} e^{-\lt(\lt[q/2+m\Delta/q\rt]^2-k_\mathrm{F}^2\rt)/2mT}}{128\pi^2S\,q/\sqrt{2m\Delta}},
\e
that is maximaized for $q=\sqrt{2m\Delta}$. The maximal value of $\delta B_{\bb{q}}$ becomes comparable to the non-quasiparticle contribution provided $T>(\Delta-\ep_\mathrm{F})/\ln{\frac{2^5\pi^3S}{3(\Delta/W)^{1/2}(\ep_\mathrm{F}/W)}}$. For EuO with $\ep_\mathrm{F}\approx 0.3$\,eV the latter condition corresponds to $T>600$\,K. The results for this regime are shown at the lower panel of Fig.~\ref{fig:Nq}.

To analyze the effect of a finite $N_{\bb{q}}$ on $\Gamma$ we take advantage of the condition $\ep_\textrm{F}\ll\Delta$, even though it is not very well fulfilled in practice.  From Eq.~(\ref{Bq}) we conclude that $N_{\bb{q}}$ varies on the scale $\delta q \sim\sqrt{m\Delta}\gg k_\mathrm{F}$ and can be replaced with a constant $N_{\bb{k}}$ in Eq.~(\ref{Bq}) within the integration range $(q_-,q_+)$. This approximation yields
\be
\label{approximation}
\Gamma(\ep,\bb{k})= \frac{\Delta^2}{2\pi S} mV_Dk_\textrm{F}\lt(1+N_{\bb{k}}-f(\ep)\rt),
\e
for $T\gg D(k+k_\textrm{F})^2$. The result of Eq.~(\ref{approximation}) overestimates the value of the integral for $q\lesssim\sqrt{m\Delta}$ and slightly underestimates it in the opposite case, but it's accurate in the limit $\ep_\textrm{F}/\Delta\ll1$. Similarly, in Eq.~(\ref{kineqsimpl}), one may replace $\bb{k}+\bb{q}$ in the integrand by $\bb{q}$ with the same accuracy. This yields the non-linear equation
\be
\label{Equation1}
dN_{\bb{q}}/dt= \big(1-N_{\bb{q}}/n_{\bb{q}}\big)\big(1+2N_{\bb{q}}\big)B_{\bb{q}},
\e
instead of Eq.~(\ref{Equation0}).  The Eq.~(\ref{Equation1}) is solved by
\be
\label{sol2}
N_{\bb{q}}=\frac{n_{\bb{q}}\lt(1-e^{-B_{\bb{q}}t(2+1/n_{\bb{q}})}\rt)}{1+2n_{\bb{q}}e^{-B_{\bb{q}}t(2+1/n_{\bb{q}})}}\approx \frac{n_{\bb{q}}\lt(1-e^{-2B_{\bb{q}}t}\rt)}{1+2n_{\bb{q}}e^{-2B_{\bb{q}}t}}.
\e
For the last approximation we used $n_{\bb{q}}\approx T/\omega_{\bb{q}}\gg1$. Unlike in the solution of Eq.~(\ref{Equation0}), the growth rate of magnon occupation in Eq.~(\ref{sol2}) increases in the entire applicability range of our theory $(N_{\bb{q}}\ll 2S)$. The characteristic time of magnon equilibration,  $\tau_q \sim \ln{(n_{\bb{q}})}/B_{\bb{q}}$, is minimal for $q\simeq\sqrt{2m\Delta/\ln{(T/2mD\Delta)}}$.

The magnon-magnon interaction is described in Debye approximation by the term \cite{Akhiezer,magnonmagnon}
\be\n
H_{mm}=-D/4S\!\!\!\!\!\s_{\bb{q}_1\bb{q}_2\bb{q}_3\bb{q}_4}\!\!\!\!\!\!b\h_{\bb{q}_1}b\h_{\bb{q}_2}b_{\bb{q}_3}b_{\bb{q}_4}(\bb{q}_1\bb{q}_2+\bb{q}_3\bb{q}_4)\delta_{\bb{q}_1+\bb{q}_2,\bb{q}_3+\bb{q}_4},
\e
which defines the magnon-magnon collision integral
\begin{align}
&\mathcal{I}_{mm}[N_{\bb{q}}]= \frac{D^2V_D^2}{4S^2}\int \frac{d^3\bb{q}_2\,d^3\bb{q}_3\,d^3\bb{q}_4}{(2\pi)^5}\;
\delta(\bb{q}+\bb{q}_2-\bb{q}_3-\bb{q}_4)\n\\
&\times  \lt[(N_{\bb{q}}\!+\!1)(N_{\bb{q}_2}\!+\!1)N_{\bb{q}_3}N_{\bb{q}_4}-N_{\bb{q}}N_{\bb{q}_2}(N_{\bb{q}_3}\!+\!1)(N_{\bb{q}_4}\!+\!1)\rt]
\n\\
&\times(\bb{q}\bb{q}_2)^2  \delta(\omega_{\bb{q}}+\omega_{\bb{q}_2}-\omega_{\bb{q}_3}-\omega_{\bb{q}_4}),
\label{mm}
\end{align}
that is formally of the same order as $\mathcal{I}_{sd}$ with respect to the small parameter $1/2S$. Thus, to be consistent we have to add $\mathcal{I}_{mm}$ to the right-hand side of Eq.~(\ref{kinetic}). We refer to the corresponding solution as $N_{\bb{q}}+\delta N_{\bb{q}}$, where $\delta N_{\bb{q}}$ is the correction due to the magnon-magnon interaction.
	
We note, however, that the magnon-magnon interaction conserves the total number of magnons since $\sum_{q} \mathcal{I}_{mm}[N_{\bb{q}}]=0$. Moreover, the effect of magnon-magnon interaction for $t \lesssim B_0^{-1}$ gives only a small correction $\delta N_{\bb{q}} =C_{\bb{q}} t^3$ to the solution of Eq.~(\ref{kinetic}). To estimate this correction we plug in $N_{\bb{q}}(t)\approx B_{\bb{q}}t$ into Eq.~(\ref{mm}). The results obtained are illustrated in the inset in Fig.~\ref{fig:Nq}.
	
From $N_{\bb{q}}$ we can easily access such experimentally relevant quantities as the unit cell magnetic moment  $M(t)= -V_D\int d^3q\,N_{\bb{q}}(t)/(2\pi)^3$ and the energy transfer per unit cell $E_m(t)=V_D\int d^3q\,N_{\bb{q}}(t)\omega_{\bb{q}}/(2\pi)^3$ as
\beml
\beq
\delta M(t) &\simeq& - \frac{3}{32\pi^2} \lt(\frac{\Delta}{W}\rt)^{3/2}\!\!\!\frac{t}{B_0^{-1}},\\
E_m(t)&\simeq&\frac{3}{8\pi^3} \lt(\frac{\Delta}{W}\rt)^2 \omega_D\frac{t}{B_0^{-1}}.
\eq
\eml
where we assume $t\lesssim B_0^{-1}$.
	
To conclude, we investigated the energy and momentum transfer from hot itinerant-electron subsystem to the localized one in half-metallic ferromagnets, including ferromagnetic semiconductors. We demonstrated that this transfer is completely determined by non-quasiparticle states, hence a naive quasi-single-particle picture is inadequate for describing this effect. Thus, we showed that subtle correlation effects are of crucial importance for non-equilibrium magnetism. Importantly, the time of energy and momentum transfer cannot be estimated from a simple dimensional analysis even to the order of magnitude. The quantitative consideration presented gives rise to a numerical factor $A_S \approx 10^5$ (for $S=7/2$) which is responsible for a strong suppression of the energy transfer rate with respect to what one could naively expect. Another important result is a relatively small role of magnon-magnon interactions in comparison to that of $s$-$d$ coupling. This is not {\em a priori} obvious since these two contributions have formally the same order of magnitude in the collision integral with respect to the only small parameter of the perturbation theory $1/2S$.
	
{\it Acknowledgments}---The authors acknowledge support from the ERC Advanced Grant 338957 FEMTO/NANO, from the Dutch Science Foundation NWO/FOM 13PR3118, from the EU Network FP7-PEOPLE-2013-IRSES Grant No 612624 ``InterNoM'' and from the NWO via the Spinoza Prize.

\supplementarystart

\centerline{\bfseries\large ONLINE SUPPLEMENTAL MATERIAL}
\vspace{6pt}
\centerline{\bfseries\large Magnon activation by hot electrons via non-quasiparticle states}
\vspace{6pt}
\centerline{S.\,Brener, B.\,Murzaliev, M.\,Titov, M.\,I.\,Katsnelson}
\begin{quote}
In this Supplemental Material we provide technical details for the analysis of the effect of magnon-magnon scattering.
\end{quote}

\section{Derivation of Eq.~(\ref{mm})}

The Heisenberg Hamiltonian for interacting localized spins is given by
\be
\label{heisenberg}
H_H=-\s_{ij}J_{ij}\bb{S}_i\bb{S}_j,
\e
where the sum runs over the lattice sites and $J_{ij}= J(\bb{r}_i-\bb{r}_j)$ denote the corresponding exchange interaction between the spins on the sites $i$ and $j$, that depends on the distance between the corresponding coordinates $\bb{r}_i$ and $\bb{r}_j$. To formulate the bosonic Hamiltonian one applies the Holstein-Primakoff transformation
\be
S_{i}^z=S-b\h_ib\0_i,\qquad S^{+}_i=\sqrt{2S}b\h_i\sqrt{1-\frac{b\h_ib\0_i}{2S}},\qquad S^{-}_i=\sqrt{2S}\sqrt{1-\frac{b\h_ib\0_i}{2S}}b\0_i,
\e
where $S$ is the absolute value of the spin while $b\h_i$ and $b\0_i$ are magnon creation and annihilation operators, correspondingly. By disregarding three and higher-order magnon processes (due to the condition $2S\gg 1$) one maps Eq.~(\ref{heisenberg}) on the bosonic model $H_H=H_d+H_{mm}$, where
\be
H_d=2S\sum_{\bb{q}}\big(J(0)-J(\bb{q})\big)b\h_{\bb{q}}b_{\bb{q}}=\sum_{\bb{q}}\omega_{\bb{q}}b\h_{\bb{q}}b_{\bb{q}}.
\e
defines the spectrum of non-interacting magnons. In the Debye approximation we assume that quadritic dispersion $\omega_{\bb{q}}=2S\big(J(0)-J(\bb{q})\big)=Dq^2$ extends up to the momentum cut off at $q=q_D$. The magnon-magnon scattering processes are, then, taken into account by the term
\be
H_{mm}=\frac{1}{2} \s_{ij}J_{ij}\lt(b^{\h}_ib^{\h}_ib_ib_j+b\h_ib\h_jb_jb_j-2b\h_ib\h_jb_jb_i\rt),
\e
where we have used that $J_{ii}=0$. The Fourier transform yields the expression
\be
H_{mm}=\frac{1}{2}\sum_{\bb{q}_1,\bb{q}_2,\bb{q}_3,\bb{q}_4}\lt(J(\bb{q}_1)+J(\bb{q}_4)-2J(\bb{q}_1-\bb{q}_4)\rt) b\h_{\bb{q}_1}b\h_{\bb{q}_2}b_{\bb{q}_3}b_{\bb{q}_4}\delta_{\bb{q}_1+\bb{q}_2,\bb{q}_3+\bb{q}_4},
\e
which is readily simplified using Debye approximation as
\be
\n
\mathcal{H}_{mm}=\frac{D}{8S}\sum_{\bb{q}_1,\bb{q}_2,\bb{q}_3,\bb{q}_4}\!\!\!\! b\h_{\bb{q}_1}b\h_{\bb{q}_2}b_{\bb{q}_3}b_{\bb{q}_4}\delta_{\bb{q}_1+\bb{q}_2,\bb{q}_3+\bb{q}_4}\lt((\bb{q}_1\!-\!\bb{q}_4)^2+(\bb{q}_1\!-\!\bb{q}_3)^2+(\bb{q}_2\!-\!\bb{q}_4)^2+(\bb{q}_2\!-\!\bb{q}_3)^2-q_1^2-q_2^2-q_3^2-q_4^2\rt).
\e
With the help of the moment conservation the latter expression can be cast in the following form
\be
\mathcal{H}_{mm}=-\frac{D}{4S}\s_{\bb{q}_1,\bb{q}_2,\bb{q}_3,\bb{q}_4} b\h_{\bb{q}_1}b\h_{\bb{q}_2}b_{\bb{q}_3}b_{\bb{q}_4}(\bb{q}_1\bb{q}_2+\bb{q}_3\bb{q}_4)\delta_{\bb{q}_1+\bb{q}_2,\bb{q}_3+\bb{q}_4},
\e
that leads to the Eq.~(\ref{mm}) from the main text by applying the Fermi's golden rule.

\section{Evaluating Eq.~(\ref{mm})}

\begin{figure}
\includegraphics[width=.45\columnwidth]{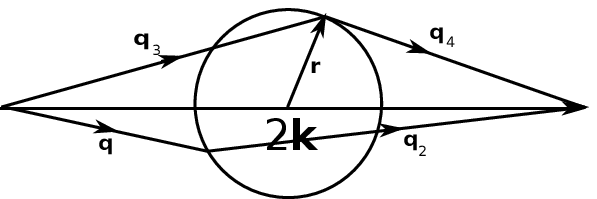}
\caption{Geometric view of magnon scattering.}
\label{fig:S1}
\end{figure}

In order to evaluate Eq.~(\ref{mm}) numerically one needs to pay attention to the momentum integration range. In the Debye model, the latter is determined by the ultra-violet cut-off $q_D$, which has to be treated with care not to violate the detailed balance of magnon scattering. It is convenient to introduce symmeterized integration variables $Q$ and $\bb{r}$ by means of the relations
\be
2Q^2=q^2+q^2_2=q_3^2+q_4^2, \qquad \bb{r}=(\bb{q}_3-\bb{q}_4)/2,
\e
and define the integration range for these variables from the condition that the initial magnon momenta do not exceeds $q_D$.  Note, that the variable $Q^2$ defines the total energy of the incoming magnons, while the vector $\bb{r}$ is sketched in Fig.~\ref{fig:S1}.
In addition we introduce an auxiliary vector $\bb{k}=(\bb{q}_3+\bb{q}_4)/2$ that defines the total momenta of the incoming magnons. Note, that $k^2+r^2=Q^2$. It is evident from Fig.~\ref{fig:S1} that the value of $\bb{q}_2$ is fixed for any given $Q^2$ and $r^2$. (The overall rotation of $\bb{k}$ around $\bb{q}$ is trivial and gives the factor $(2\pi)^2$.) In order to fix $\bb{q}_{3}$ and $\bb{q}_{4}$ we also need to know the angle between $\bb{r}$ and $\bb{k}$.

In new variables it becomes straightforward to get rid of $\delta$-functions. This results in the following transformation
\be
\int_0^{q_D} \frac{d^3\bb{q}_2\,d^3\bb{q}_3\,d^3\bb{q}_4}{(2\pi)^5}
\delta(\bb{q}+\bb{q}_2-\bb{q}_3-\bb{q}_4)\delta(q^2+q^2_2-q^2_3-q^2_4)= \frac{1}{q}\int_{q^2/2}^{q^2/2+q_D^2/2}d(Q^2)\int \frac{r^2drd(\cos{(\widehat{\bb{r},\bb{k}})})}{(2\pi)^3}.
\e
where the integration limits for $\bb{r}$ are taken from the condition that $\bb{q}$, $\bb{q}_2$ and $2\bb{k}$ form a triangle. This implies the condition
\be
k^2\in(\frac{(q-q_2)^2}{4};\frac{(q+q_2)^2}{4}).
\e

Since $q_2^2=2Q^2-q^2$ and $k^2=Q^2-r^2$ we also conclude
\be
\label{r1}
r^2\in\lt(\frac{Q^2-q\sqrt{2Q^2-q^2}}{2};\frac{Q^2+q\sqrt{2Q^2-q^2}}{2}\rt),
\e
which means that the integrating limits for $r^2$ do not depend on the ultra-violet cut-off. The presence of a finite cut-off provides yet additional restrictions. To find them
we need to make sure that $q_{3,4}^2<q_D^2$. Introducing $x=\cos{(\widehat{\bb{r},\bb{k}})}$ we write
\be
q_{3,4}^2=Q^2\pm2x\sqrt{r^2(Q^2-r^2)}<q^2_D,
\e
that provides us an additional condition for $x$ in the form
\be
\label{xcond}
|x|<\frac{\Lambda^2-Q^2}{2r\sqrt{Q^2-r^2}}.
\e
This condition is restrictive provided
\be
\frac{\Lambda^2-Q^2}{2r\sqrt{Q^2-r^2}}<1
\e
otherwise the integration limits for $x$ are merely $(-1,1)$. The latter is true for
\be
\label{r2}
r^2\in\lt(\frac{Q^2-q_D\sqrt{2Q^2-q_D^2}}{2};\frac{Q^2+q_D\sqrt{2Q^2-q_D^2}}{2}\rt),
\e
assuming $Q^2>q_D^2/2$ but not otherwise. Note that the condition (\ref{r2}) is always a subset of the condition (\ref{r1}) as fas as $q<q_D$.
	
Collecting Eqs.~(\ref{r1}), (\ref{xcond}), and (\ref{r2}), normalizing the momenta by $\sqrt{2m\Delta}$, and introducing $\Lambda^2=q_D^2/2m\Delta=W/\Delta$ we can symbolically write the integration range in the following way:
\begin{align}
&\frac{1}{2q}\int_{q^2/2}^{\Lambda^2/2}d(Q^2)\int_{\frac{Q^2-q\sqrt{2Q^2-q^2}}{2}}^{\frac{Q^2+q\sqrt{2Q^2-q^2}}{2}}rd(r^2)\int_{-1}^1dx\\
&
+\frac{1}{2q}\int_{\Lambda^2/2}^{\Lambda^2/2+q^2/2}d(Q^2)\lt[\lt(\int_{\frac{Q^2-q\sqrt{2Q^2-q^2}}{2}}^{\frac{Q^2-\Lambda\sqrt{2Q^2-\Lambda^2}}{2}}+\int^{\frac{Q^2+q\sqrt{2Q^2-q^2}}{2}}_{\frac{Q^2+\Lambda\sqrt{2Q^2-\Lambda^2}}{2}}\rt)rd(r^2)\int_{-1}^1dx+\int^{\frac{Q^2+\Lambda\sqrt{2Q^2-\Lambda^2}}{2}}_{\frac{Q^2-\Lambda\sqrt{2Q^2-\Lambda^2}}{2}}rd(r^2)\int_{-\frac{\Lambda^2-Q^2}{2r\sqrt{Q^2-r^2}}}^{\frac{\Lambda^2-Q^2}{2r\sqrt{Q^2-r^2}}}dx\rt].\n
\end{align}
The integrand is given by
\be
\frac{2\omega_D}{A_S\Lambda^8}(Q^2-2r^2)^2F\lt(q^2,2Q^2-q^2,Q^2+2x\sqrt{r^2(Q^2-r^2)},Q^2-2x\sqrt{r^2(Q^2-r^2)}\rt),
\e
where $F(q^2,q_2^2,q_3^2,q_4^2)=(N_{\bb{q}}\!+\!1)(N_{\bb{q}_2}\!+\!1)N_{\bb{q}_3}N_{\bb{q}_4}-N_{\bb{q}}N_{\bb{q}_2}(N_{\bb{q}_3}\!+\!1)(N_{\bb{q}_4}\!+\!1)$.

\end{document}